\documentclass[]{spie}  

\usepackage{amsmath,amsfonts,amssymb}
\usepackage{graphicx}
\usepackage[colorlinks=true, allcolors=blue]{hyperref}
\usepackage{caption}
\usepackage{subcaption}
\usepackage{gensymb}
\usepackage{xcolor}

\usepackage{soul}

\usepackage{siunitx}
\newcommand{\SIadj}[2]{\SI[number-unit-product={\text{-}}]{#1}{#2}}

\makeatletter
\newcommand{\Rustam}{\@ifstar{\@Rb}{\@Ra}}
\newcommand{\@Rb}[1]{\textcolor[HTML]{0da34e}{#1}}
\newcommand{\@Ra}[1]{\textcolor[HTML]{0da34e}{\textbf{Rustam:} #1}}

\title{Vector beam mapping at millimeter wavelengths using a robot arm}

\author[a]{Rustam Balafendiev}
\author[a]{Thomas Gascard}
\author[a,b]{Jon E. Gudmundsson}
\affil[a]{Science Institute, University of Iceland, 107 Reykjavik, Iceland}
\affil[b]{The Oskar Klein Centre, Department of Physics, Stockholm University, AlbaNova, SE-10691 Stockholm, Sweden}

\authorinfo{Further author information: (Send correspondence to R.B. and J.E.G.)\\R.B.: E-mail: rub8-at-hi-dot-is, Telephone: +354 765 5316\\J.E.G.: E-mail: jegudmunds-at-hi-dot-is, Telephone +354 525 4625}

\begin{document} 
\maketitle


\begin{abstract}
Many experimental efforts are striving to provide deep maps of the cosmic microwave background (CMB) to shed light on key questions in modern cosmology. The primary science goal for some of these experiments is to further constrain the energy scale of cosmic inflation. It has been shown that these experiments are particularly sensitive to optical systematics. Near-field vector beam mapping, or holography, is now employed in a variety of CMB-focused experimental efforts due to the technique's ability to provide full details of electromagnetic field propagation through complex systems. In this proceeding, we describe the development of a measurement bench for millimeter-wave phase-sensitive beam mapping with the goal of characterizing optical components for CMB experiments. We discuss the testing of a beam scanner based on a 6-axis robot arm, the related custom control software, the readout architecture, and the overall validation of the system through various testing procedures. Dynamic range of \SI{70}{\deci\bel} is demonstrated for the presented setup. With the current mechanical setup, we derive an upper limits of \SI{45}{\micro\meter} on the absolute positioning error and \SI{10}{\micro\meter} on positional repeatability.
\end{abstract}                  

\section{INTRODUCTION}
\label{sec:intro}  


Measurements of the cosmic microwave background (CMB) that meet the science goals of upcoming experiments call for strict requirements on various instrument performance metrics.\cite{CMBS42017} This is particularly true for experiments that search for primordial B-mode polarisation,\cite{Kamionkowski16} where a detailed understanding of optical performance is critical to success.\cite{Hanson2010P} For CMB experiments in particular, coherent mm-wave measurements of both optical components and full telescope assemblies can provide valuable data to inform instrument models used for cosmological analysis. This technique is currently employed, both at room and cryogenic temperatures, to advance the understanding of the optical performance of instruments currently in development.\cite{Chesmore22, Nakano23, Takakura23} In this paper we describe a system for coherent mm-wave measurements which uses a commercial 6-axis industrial robot to perform the scanning instead of a system composed of multiple linear and rotation stages.

Robot arms have already been used for phase-sensitive microwave measurements \cite{Novotny17,Matos19,Boehm17}. The use of robot arms offers some advantages over more traditional linear translation methods frequently used for the characterization of optical components \cite{Chesmore22,Gascard:23}. This includes ease of calibration and improved workspace flexibility. In this paper we describe such a system, demonstrate its usefulness for a measurement of a far field of a horn antenna, specifically stressing the ease of calibrating for cross-polarisation, and attempt to benchmark the positional accuracy of the robot arm using the measured phase data. We also present custom open source software that can both be used to perform similar measurements in the future and be adapted for other measurement setups.







\begin{figure}[pht]
    \centering
    \includegraphics[width=.5\textwidth]{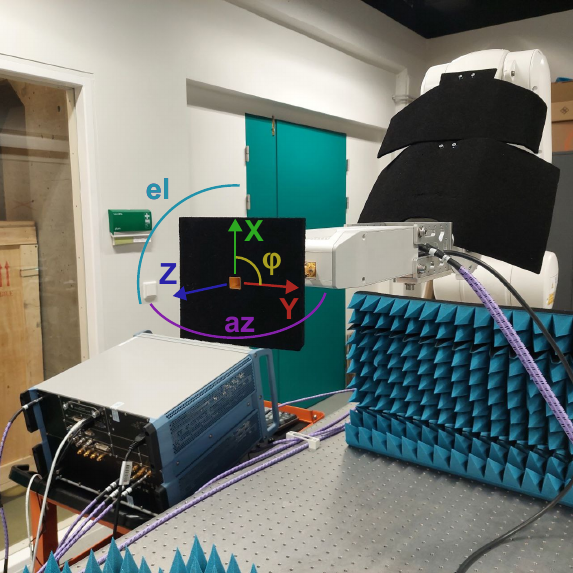}
    \caption{The 6-axis Kuka robot with a ZRX110 W-band reciever attached to a custom aluminum mounting plate that bolts to the flange of the robot. The Cartesian directions of movement are marked near the aperture of the horn, with Z being the aperture normal. Angle $\phi$, positioned in the XY plane and marked in yellow is used for polarisation modulation. Azimuth (YZ, magenta) and elevation (XZ, cyan) are used to characterise the far field beam profiles of the receiver horn.}
    \label{fig:robot}
\end{figure}

\section{System parameters and the control software}
\label{sec:system}

The system presented in this paper consists of a R\&S ZNA26A Vector Network Analyzer (VNA), W-band (75-\SI{110}{\giga\hertz}) ZC110 and ZRX110 frequency extension modules from R\&S, and a KUKA KR 6 R900-2 6-axis robot arm (Fig.~\ref{fig:robot}). The ZRX110 receiver is attached to the robot arm flange and aligned along the optical axis of the ZC110 transceiver placed approximately \SI{1.8}{\meter} away. The position of the center of the horn aperture is used as the initial Tool Center Point (TCP) within the RoboDK\cite{RoboDK} software, a simulation environment used to preemptively visualise the planned scan and send simple movement commands in its Python API to the robot controller. A filter bandwidth of \SI{1}{\kilo\hertz} and no averaging are used in the VNA data processing.

\begin{figure}[pht]
    \centering
    \includegraphics[width=1.0\textwidth]{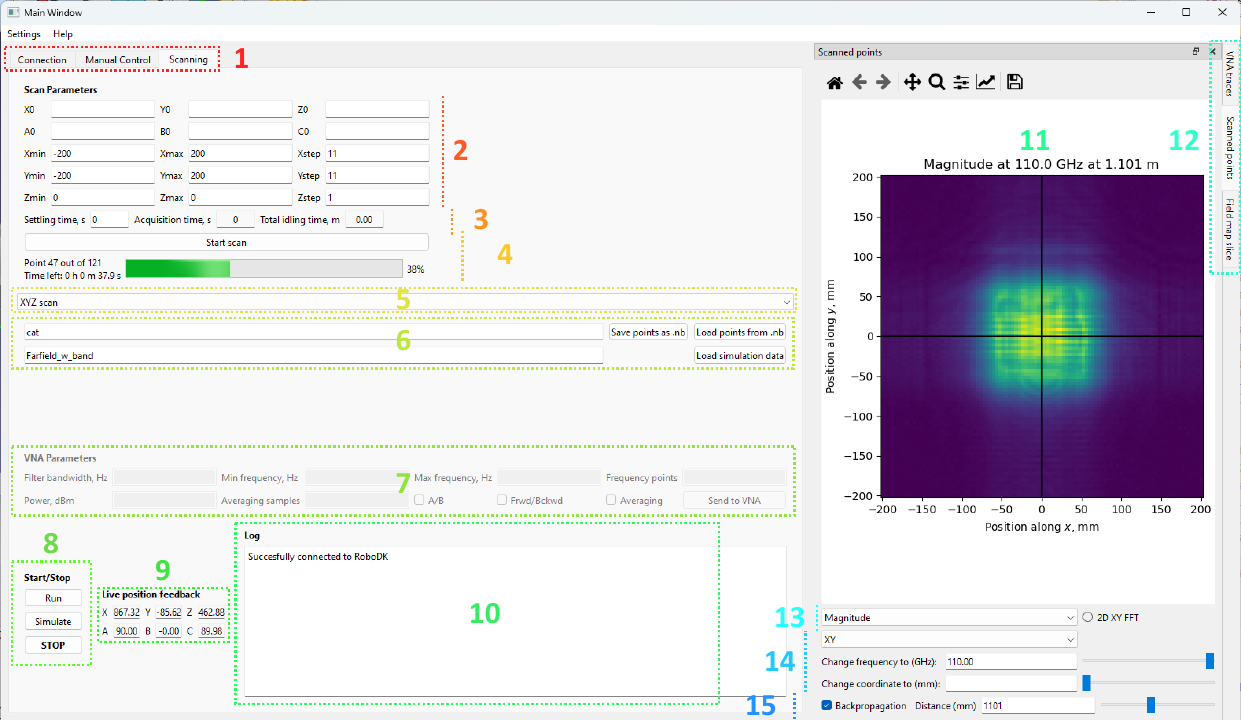}
    \caption{An example of how the user interface of the developed software looks in the process of taking data. A detailed description of the numbered sections is presesnted in section \ref{sec:system}.}
    \label{fig:ui}
\end{figure}

Manual and automated movements of the robot and parameters for the frequency sweep are handled and passed through to the VNA and robot controller via a custom software programmed in Python and available in the following \href{https://github.com/Skuggsja-Lab/skuggsja-scan}{GitHub repository}. The control software further offers a PyQT-based GUI overlay, simplifying the manual control of the robot and the scan setup. Presently, two choices of scanning trajectories are implemented: a scan in the Cartesian XYZ coordinates relative to the TCP and an angular scan which sweeps the receiver horn along elevation and azimuth around the TCP. The starting position of the scan and orientation of the horn can be manually adjusted. The software also communicates with the VNA in order to synchronize the movements of the robot with the data acquisition and to obtain the raw complex-valued frequency spectrum at the scanned points, which is then written to a 4D ($x,y,z,f$) Numpy array.\cite{Numpy} The GUI also allows for live monitoring of the scan via a 2D colormesh plot and 1D slices, comparing the experimental data with ones obtained numerically and post-processing of the data directly in the software. Currently, the post-processing capabilities include a 2D Fourier transform of the scan and back-propagation of the obtained vector fields along the Z-direction (see Figure~\ref{fig:robot}). An example of the user interface in the process of taking data is shown in Figure~\ref{fig:ui} with various sections labeled. The workflow is divided into three main window tabs (1): the first one is used to establish the connection with the VNA and the robot, the second allows for manual control of the robot position as well as polarisation modulation and the third one houses the scan settings. Various fields at the top of the scanning tab are (2) used to set the initial point, orientation, span and resolution of the scanned volume or plane. The section right after (3) houses the field used for setting the settling time, an initial estimation of the scan duration and the "Start scan" button. 

An accurate estimation of the scan duration, visualised with a progress bar (4), is made by extrapolating the average length of time required to move to and measure a single point. Below is the drop-down menu for selection of different scanning modes (5) and the buttons (6) for saving and loading the measured data as well as loading the simulated results for comparison. The lower half of the main window houses the fields for changing the VNA settings (7), buttons for switching between running the scan in the software and running it on the physical robot (8), live position monitoring (9) and the message log (10). The detachable windows on the right display various plots (11) of the measured data, depending on the chosen tab (12). Currently the three options are: live monitoring of the traces from the VNA, a colormap of the scanned surface and a 1D slice of the same surface through the user-selected slice. The colormap tab presented in the figure has a few controls at the bottom. First is the drop-down menu (13) for selecting the format of the data: magnitude (linear or dB), phase, real and imaginary parts. A toggle for calculating the Fourier image of the data shown is positioned to the right. Next are various elements for controlling the displayed slice of the 4D ($x,y,z,f$) array (14). And last is the toggle for propagating the measured field backward by the distance set by the slider (15). Various settings, such as VNA and scan parameters can be saved and loaded using .toml files.

\begin{figure}[t!]
    \centering
    \includegraphics[width=0.8\textwidth]{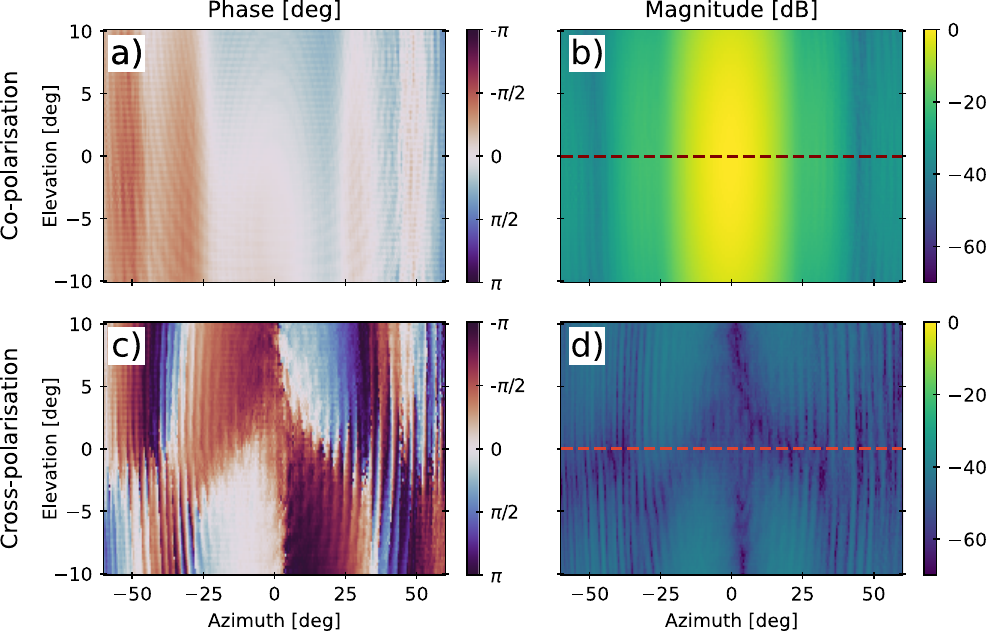}
    \caption{Angular far field scan results at \SI{75}{\giga\hertz} for the co- (top) and cross-polarisation (bottom) cases. a) Phase distribution of co-polarisation. Some skew is noticeable due to fault in alignment, but overall the phase variation remains within a \ang{145} range over the \ang{120} of azimuth that are covered by the scan. b) Normalised magnitude distribution of co-polarisation, in dB. A slice at \ang{0} of elevation is shown in dark red in Fig.~\ref{fig:slice}. c) Phase distribution of cross-polarisation. While there is some warping, the characteristic cross-polarisation pattern of the phase having discontinuities at \ang{0} of elevation and \ang{0} of azimuth can be observed. b) Normalised magnitude distribution of cross-polarisation, in dB. A slice at \ang{0} of elevation is shown in light red in Fig.~\ref{fig:slice}.}
    \label{fig:scan}
\end{figure}

\section{Results}
\subsection{Dynamic range, far sidelobes and polarization modulation}
In order to determine the dynamic range of the system we have performed a scan within a range of angles in the Ludwig 2 azimuth-over-elevation coordinate system\cite{ludwig_pol}. Filter bandwidth of \SI{10}{\kilo\hertz} and no  averaging were used during the measurement. Each frequency sweep was comprised of 201 samples spanning the entire W-band (75-\SI{110}{\giga\hertz}), taking about 0.02~seconds to complete. Angles in the range of \ang{-60} to \ang{60} along the azimuth and \ang{-10} to \ang{10} along elevation were scanned by rotating the horn around its phase center. Within RoboDK this is represented by the position target being rotated around its X- and Y-axes respectively. The optimal position of the phase center relative to the horn aperture, which matches the position of the tool center point in RoboDK software, is found by repeating this scan while varying the position of the rotation center and minimising the measured change in phase. A 6-axis robot enables the required rotations without the need of additional mechanical systems. By developing software that enables synchronization between the robot control and the signal readout, we enable automated calibration routines which will be the subject of future work. 



\begin{figure}[hpt]
    \centering
    \includegraphics[width=\textwidth]{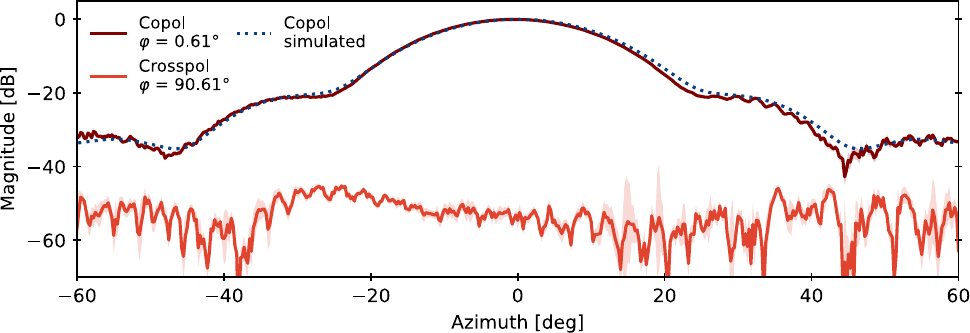}
    \caption{Comparison of \ang{0} of elevation slices for the cross-polarisation measurement and co-polarisation measurement and simulation. A 3$\sigma$ deviation envelope obtained from 40 identical measurements is plotted over the experimental data. The simulated data has been obtained using a model of the horn in CST MWS.\cite{CST} A rotation \ang{.61} was added to $\phi$ in the process of minimising the cross-polarisation response. The measured result matches well with the simulated one, with notable discrepancies being due to standing waves between the transmitter and the receiver and some degree of non-ideality of alignment. The normalised cross-polarisation measurement averages at about \SI{-54}{\deci\bel}, with some points going as low as \SI{-70}{\deci\bel}. This suggests a comparable dynamic range of about \SI{-70}{\deci\bel} for this system.}
    \label{fig:slice}
\end{figure}

The same scan was then repeated with the horn rotated \ang{90} around the $Z$-axis in order to obtain the cross-polarisation far-field characteristic of the receiver horn. The addition of \ang{.61} on top of the \ang{90} rotation on $\phi$, directly controlled by the flange axis of the robot, was informed by varying $\phi$ to minimize the cross-polarisation signal when both the receiver and the transmitter are supposedly co-linear, thus excluding the co-polarisation contribution to the measured cross-polarisation signal. As can be seen in Figures~\ref{fig:scan}b and \ref{fig:slice}, the measured co-polarization beam cut closely matches the simulated beam down to the \SIadj{40}{\deci\bel} range, barring some standing wave contributions visible at the sidelobes. The maximum of the co-polarisation beam is a few degrees off-center. This, combined with the slant of the phase distribution visible in Fig.~\ref{fig:scan}a, suggests that these result can be further improved with a better alignment. The cross-polarisation measurement suggests that the dynamic range of the system is around \SI{70}{\deci\bel} (see Figures \ref{fig:scan}d and \ref{fig:slice}). In order to verify the robustness of this result, 40 more identical measurements were performed. The obtained 3$\sigma$ deviation margins are plotted as semi-transparent regions for both co- and cross-polarisation in Figure~\ref{fig:slice}. Their low value at several points around the \SI{-70}{\deci\bel} mark supports this evaluation of the dynamic range.

\subsection{Positional repeatability}

Traditionally, the positional accuracy and repeatability of a 6-axis robot can be evaluated using a laser tracker or a set of micrometre dial gauges.\cite{Novotny13} Since the robot is equipped with a phase-sensitive mm-wave receiver, we can instead use the data that we obtain with the receiver to extract robot positional information with high accuracy.  


Generally, the relationship between the position of the receiver and the measured phase of a wavefront $Ae^{i\theta} = Ae^{i(\textbf{kr}+\omega t)}$ propagating through free space can be expressed as:
\begin{equation}
    |\textbf{r}| = \frac{\theta}{|\textbf{k}|} = \frac{\theta}{k_0}  = \frac{\theta c}{2 \pi f},
\end{equation}
where $\textbf{r}$ is the spatial coordinate, $\textbf{k}$ is the wave vector, $k_0$ is the wavenumber in free space, $\theta$ is the phase and $f$ is the frequency. If the receiver is positioned in the far field and is aligned with the transmitter on the Z-axis, then the measured wavefront can be approximated as a plane wave $Ae^{i(k_zz+\omega t)}$. As long as this is the case, the position of the receiver can be found as:
\begin{equation}
    z = \frac{\theta}{k_z} = \frac{\theta c}{2 \pi f}.
\end{equation}
The following discussion assumes the initial position of the robot to be at $z$ = \SI{0}{\milli\meter}. By keeping track of phase information, we see that positional errors stay well within one wavelength. We note that this method does not offer a way to distinguish between the change of phase stemming from a change in position and a change of phase stemming from unknown sources. 



In order to test the positional repeatability of the robot arm we have run a series of linear scans along the Z-direction, repeating measurements at 80~locations linearly sampled along a \SIadj{8}{\milli\meter} span. These measurements were repeated 200~times over a period of about 4~hours. Due to the inertia of the robot arm, there is some initial vibration that the flange experiences after reaching the programmed point. The amplitude of this vibration is stated to be within \SI{300}{\micro\meter} after \SI{0.1}{\second}. To minimise its impact, a settling time of \SI{1.0}{\second} has been set for the scan. When a settling time of \SI{0.5}{\second} is used, the first 20 measurements need to be discarded due to the robot arm having much lower positional accuracy while warming up. This is consistent with statements made in the robot user manual. Future work will look into optimization of the scan parameters to maximize mapping speed.



\begin{figure}[hpt]
     \centering
     \includegraphics[width=\textwidth]{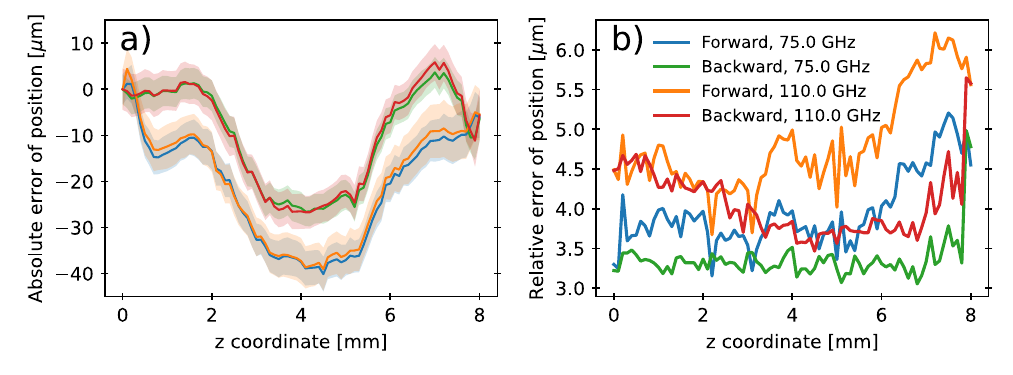}
     \caption{Position error as a function of coordinate for two directions of movement measured at the highest and the lowest frequencies. a) Absolute positional error with the semitransparent regions marking the $1\sigma$ standard deviation envelope. b) Positional repeatability as calculated by the $1\sigma$ standard deviation from 100 samples at each of the 80 locations samples along the \SIadj{8}{\milli\meter} range.}
     \label{fig:err}
\end{figure}

Since the frequency is known, phase information obtained from these scans can then be converted to an offset relative to the initial position. The absolute positional accuracy has been calculated by converting phase data to positional offsets after detrending the phase timestreams with a linear fit to minimize the effect of the phase drift, a contribution to the phase that affects all measurements and slowly changes over time, such as an ambient change in temperature or humidity.
The result of this is presented in Fig.~\ref{fig:err}a. The relative positioning error was calculated as root mean square of these values. Its values across coordinates are presented in Fig.~\ref{fig:err}b as well as shown in Fig.~\ref{fig:err}a as semitransparent regions near the traces. If the phase at the first point is assumed to correspond to $z = \SI{0}{\milli\meter}$, the absolute positioning error for this particular case seems to be at most about \SI{40}{\micro\meter}, with the positional repeatability being around \SI{4}{\micro\meter} on average. While this falls well within the bounds of \SIadj{200}{\micro\meter} absolute pose accuracy and \SIadj{20}{\micro\meter} pose repeatability claimed on the KUKA robot datasheet\cite{datasheet}, it only covers a very narrow use case (single linear motion, extremely small steps) and does not necessarily demonstrate that these values can be assumed for the entire range of motion of the robot.

\begin{figure}[]
     \centering
     \includegraphics[width=0.6\textwidth]{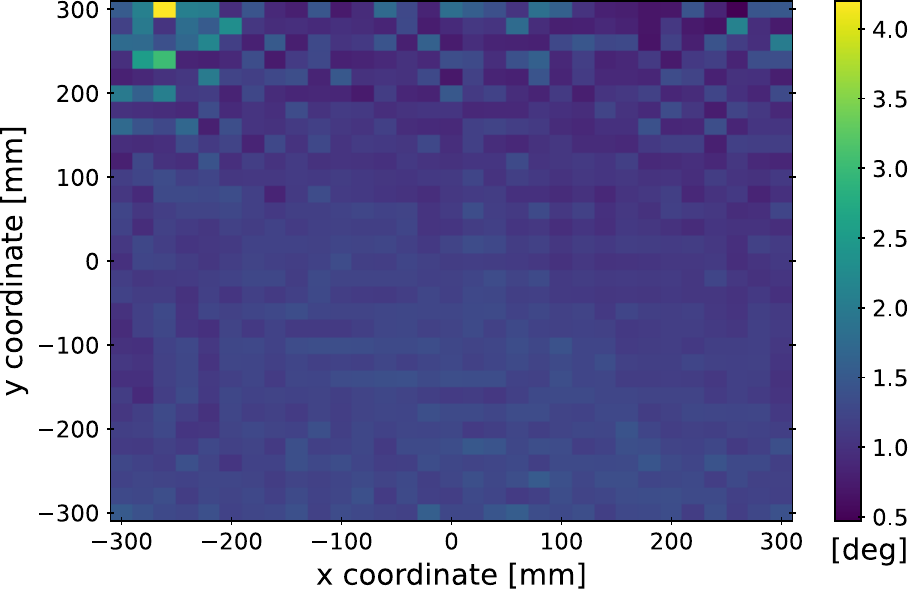}
     \caption{Positional repeatability of a scan over a $31\times31$ point grid with a span of $60\times\SI{60}{\centi\meter}$ at the lowest frequency of \SI{75}{\giga\hertz}, calculated by the $1\sigma$ standard deviation from 50 samples at each point.}
     \label{fig:err_xy}
\end{figure}

We also performed a test at \SI{75}{\giga\hertz} with the robot scanning through a $31\times31$ point grid in the XY plane with a span of $60 \times \SI{60}{\centi\meter}$. Each scan lasted for 33 minutes and was repeated 100 times for an overall measurement duration of approximately 55 hours. Since statistical errors that are not related to XYZ position errors, including tip and tilt pointing errors, will contribute to the overall phase error that is measured at each location, the following result represents an upper bound on the relative error of position. The phase error calculated across this grid is shown in Fig.~\ref{fig:err_xy}. The maximum $1\sigma$ error is approximately \ang{4.26}, which suggests an upper limit of \SI{47}{\micro\meter} in relative position error. The mean value across the array is \ang{1.05} which corresponds to about \SI{11.7}{\micro\meter}. No phase drift correction was implemented in generating these results and measurements were conducted in a typical laboratory setting with multiple candidates for spurious reflections. 


\section{Conclusions and discussion}
\label{sec:conclusions}

We have presented a phase-sensitive microwave measurement setup involving a 6-axis robot and demonstrated its capabilities by mapping the beam of a W-band horn antenna. 
The preliminary results shown can be improved by stricter alignment of the optical components and better control of the standing waves in the system. This, as well as further improvements to the software through better scanning strategies, automated calibration routines, optimised scanning parameters and new post-processing options, will be explored in future efforts.

An open-source control software with a graphical user interface which allows for controlling the robot and synchronizing with the vector network analyzer for data acquisition has been developed and made available in a public \href{https://github.com/Skuggsja-Lab/skuggsja-scan}{repository}. The software also supports typical data processing and visualization options, including field propagation and data slicing.

Wide-angle far field measurements have been performed with the cross-polarisation results demonstrating \SI{70}{\deci\bel} of accessible dynamic range in the context of our laboratory setup. Taking \SI{350}{GHz} as a reference - the higher limit of a frequency band of high interest for CMB observations, as a reference \cite{CMBS42017} - the positional accuracy has an upper limit at \SI{45}{\micro\meter}, or 6\% wavelength, whilst the positional repeatability sits within \SI{10}{\micro\meter}, or 1\% wavelength.

Relying on the high accuracy, repeatability and versatility the 6-axis robot arm offers, the holography measurement bed discussed in this work presents an attractive alternative to a setup relying on traditional stage. This scanner can carry precise beam holography of optical components for upcoming CMB instruments.

\acknowledgments
Funded by the European Union (ERC, CMBeam, 101040169). Views and opinions expressed are however those of the author(s) only and do not necessarily reflect those of the European Union or the European Research Council Executive Agency. Neither the European Union nor the granting authority can be held responsible for them. We acknowledge support from The Icelandic Research Fund (Grant number: 2410656-051), the Swedish Research Council (Reg.\ no.\ 2019-03959), and the Swedish National Space Agency (SNSA/Rymdstyrelsen).
\bibliography{report} 
\bibliographystyle{spiebib} 

\end{document}